# Mobile Technology: A Panacea to Food Insecurity in Nigeria

Mudathir Muhammad Salahudeen[1], Muhammad Auwal Mukhtar[1], Saadu Salihu Abubakar[1]
Salawu, I.S[2]

[1]Usman Dan Fodio University, Sokoto
[2]Creative Associates International, NY – Abuja.

**Corresponding Author**

**Mudathir Muhammad Salahudeen**
Department of Computer Science,
Usmanu Danfodiyo University Sokoto,
SOKOTO
email: mudasalahu01@gmail.com
Phone No.: 09066642998
ORCID:

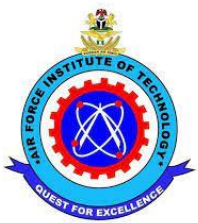

**Abstract**

*Over time, agriculture is the most consistent activity, and it evolves every day. It contributes to a vast majority of the Gross Domestic Product (GDP) of Nigeria but as ironic as it may be, there is still hunger in significant parts of the country due to low productivity in the agricultural sector and comparison to the geometric population growth. During the first half of 2022, agriculture contributed about 23% of the country's GDP while the industry and services sector had a share of the remaining 77%. This showed that with the high rate of agricultural activities, Nigeria has not achieved food security for the teeming population. and more productivity levels can be attained. Technology can/will assist Nigeria in overcoming global poverty and hunger quicker in both rural and urban areas. Today, there are many types of agricultural technologies available for farmers all over the world to increase productivity. Major technological advancements include indoor vertical farming, automation, robotics, livestock technology, modern greenhouse practices, precision agriculture, artificial intelligence, and blockchain. Mobile phones have one of the highest adoption rates of technologies developed within the last century. Digitalization will bring consumers and farmers closer together to access the shortest supply chain possible and reduce rural poverty and hunger. The paper will review the different agricultural technologies and propose a mobile solution, code Sell Harvest, to make farming more sustainable and secure food*

## 1. Introduction

In Nigeria, agriculture is one of the foundations of the Nigerian economy. It is the non-oil sector that contributes greatly to the economic growth of the country. In the first half of 2022, agriculture contributed about 23% of the country's GDP. But sadly, there have been so many factors hindering the success of agriculture in the country. More than 80 per cent of farmers in Nigeria are small-holder farmers. These farmers face many difficulties which affect their production. Smallholder farmers in Nigeria have limited access to credit facilities which limits their productivity to a great extent. Despite the fact Nigeria has a lot of cultivable lands, a great percentage of it is being converted to other uses than agriculture. Farmers lack agricultural information, and this is a factor that promotes agricultural ignorance of modern farm technologies for the value chains. There is very limited access to modern improved technologies and their general circumstance do not always merit tangible investments in capital inputs and labour. This paper will review the different agricultural technologies and propose a mobile e-commerce solution, Sell Harvest, that will improve marketing in the agricultural value chain. This will reduce postharvest losses in both urban, semi-urban, and rural areas.

## 2. Agricultural Technologies
### 2.1 Smart Farming

Smart farming refers to the use of on-farm and remote sensors to generate and transmit data about a specific crop, animal or practice to enable the mechanization and automation of on-farm practices and achieve more efficient, high-quality and sustainable production of agricultural goods. Smart farming solutions often rely on connectivity between Internet of Things (IoT)-enabled devices to optimize production processes and growth conditions while minimizing costs and optimizing resource use.

Smart farming is a method of farming based on previous inspection and research in order to make farms and fields more efficient and optimized. Smart farming solutions can play an important role in helping smallholder farmers in low- and middle-income countries (LMICs) like Nigeria increase their productivity and resilience to disaster by opening access to assets and mechanization, optimizing the use of inputs, labor, and natural resources, and reducing crop and animal losses and waste. Although smart farming is one of the more recent digital agriculture use cases to emerge in LMICs, early results appear promising. Some methods used in smart farming include Precision Farming, Automated Irrigation, Vertical Farming, and Hydroponics.

### 2.2 Livestock Farming Technology
Livestock farming is an innovative approach to raising animals that use modern technologies to gather data about every animal on a farm and use that data to optimize the management practices by either reducing inputs or increasing the overall farm productivity.

### 2.3 Agricultural Drone
This is the process of using an unmanned aerial vehicle used in agriculture operations, mostly in yield optimization and monitoring crop growth and production. Agricultural drones provide information on crop growth stages, crop health, and soil variations. Multispectral sensors are used on agricultural drones to image electromagnetic radiation beyond the visible spectrum, including near-infrared and short-wave infrared. Figure 2 shows an agricultural drone spraying a farmland

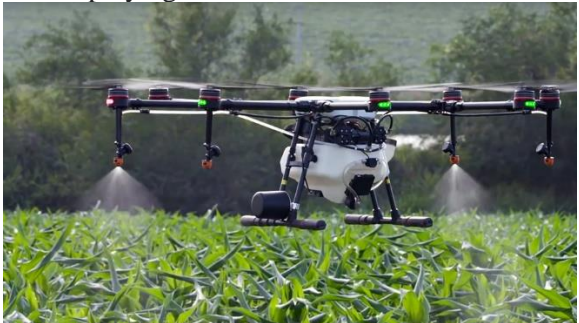

**Figure 1.** Agricultural Drone

### 3. Proposed Mobile Solution - Sell Harvest
Nigeria contributes about 51 per cent of the total food supply in West Africa but the country's post-harvest losses have increased to an estimated N3.5tn annually. Post-harvest losses in fruits and vegetables only in Nigeria are about 50 per cent annually. Talking about fruits and vegetables, rotten fruits are a very common sight at Mile 12 Market in Lagos. But it is even worse in rural areas where huge quantities of fruits and vegetables get rotten or withered badly before they get to the market. In fact, in Nigeria, if a fruit is not in season, you cannot get it to buy.

Sell Harvest is a mobile solution to post-harvest loss in Nigeria. It is a decade Agrotech Solution optimized to connect rural and smallholder farmers in Nigeria directly to potential buyers in urban areas eliminating the middlemen (main price hiking agents) and providing real-time information on market trends and prices, weather forecast, learning page, and so much more.

Paving way for innovations in the Agric sector get to the main components in the Agric chain i.e. Farmers, it also gets to bring about solutions to the farmers in terms of their farming needs and wants. The likes of consultancy, mentorship, censorship, and most especially their customers.

### 3.1 Features of Sell Harvest
#### 3.1.1 Language Localization
Language Localization can be defined as the process of adapting a product or service according to the native language and culture of a particular region. Sell Harvest is built to support translation to the 3 most dominant local languages in Nigeria, Hausa, Yoruba, and Igbo for easy use and adoption by unschooled farmers.

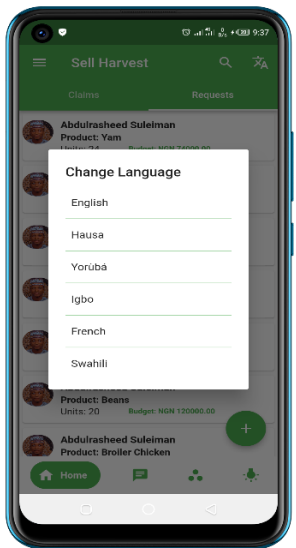

**Figure 2.** Sell Harvest Language Page

### 3.1.2 Negotiation between Buyer and Seller (Call and Chat)

To retain the ability to negotiate prices unlike other e-commerce platforms, buyers can call the farmers directly to negotiate prices when the products are to be purchased in bulk. This gives the buyer the flexibility and assurance of not being in a business transaction with a bot.

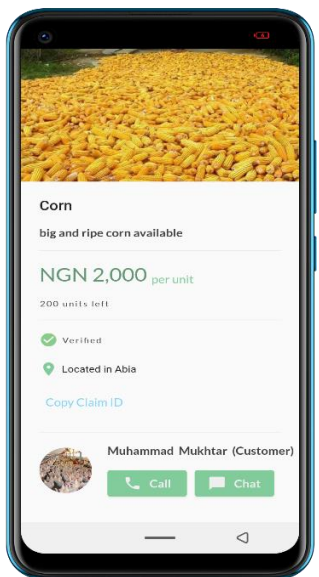

**Figure 3.** Sell Harvest Call and chat section

### 3.1.3 Verification of Users and Commodities

With the Security Risk Management Strategy, the Verification of users on the platform has been carried out by the components of the platform called Agents, and also transactions are closely monitored by the human factor too.

Commodity purchases are thoroughly checked and verified on-site by Sell Harvest Agents before the conclusion of the transaction to ensure buyers get exactly what they ordered.

### 3.1.4 Market Price Index

One of the main issues with the Agricultural market sector is that there is a lack of reliable information on the Food/Crop supply and demand, most importantly: maize, rice, beans, pepper, ginger, wheat, and soybeans (Major Crops).

Sell Harvest gives the latest market prices of these products.

### 3.1.5 Learning Page

To enable rural farmers to tap into modern ways of farming, Sell Harvest has an online educational platform where video content is uploaded for farmers to learn the trending, most recent, and efficient ways of farming, storing, and processing their produce, these video contents are made available English and the local languages, Hausa, Yoruba, and Igbo.

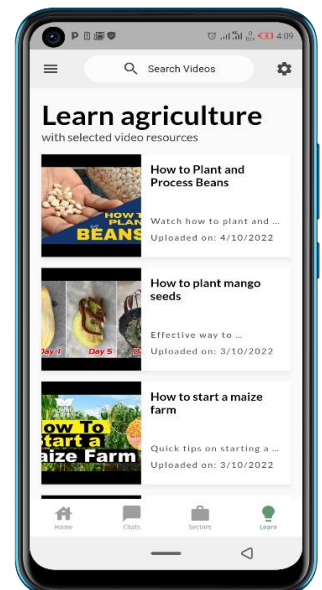

**Figure 4.** Sell Harvest Language Page

### 3.1.6 Weather Forecast

As the sacred business of Agriculture evolves around the weather and climatic factors of our ecosystem, the integration of the weather forecast system will help in making futuristic plans and decisions that will be tailored to the reality of the available data.

### 3.1.7 Sectors

As a one-stop Agrotech setup, Sectors for agro-specific Business/Services are made available, ranging from Marketing, Transportation, Renting/Leasing, Retailing, Processing, and most importantly Loan And Investment.

## 4. Conclusion

In conclusion, the solution provided by Sell Harvest will aid in eliminating post-harvest loss which always serves as a major threat to the farmer, while consumers get to purchase all the available goods from the farmers, coupled with the fact that cheaper food been derived while giving less for more (compared to the previous middlemen infested system).

## 5. References

1. Regina Nneamaka Mgbenka, Evangeline Nwakaego Mbah, Ezeano Caleb Ike 2015, A Review of Smallholder Farming in Nigeria: Need for Transformation. Agricultural Engineering Research Journal 5(2): 19-26, 2015.
2. Bamiduro, J.A. and A.G. Rotimi, 2011. Small-scale farming and agricultural product marketing for sustainable poverty alleviation in Nigeria. Open Journal System. Home, vol. 7(3).
3. Mgbenka, R.N. and A.E. Agwu, 2011. Communication platforms existing among research, extension, and farmers in Abia and Enugu States of Nigeria. A pre-Ph. D research seminar, Department of Agricultural Extension, University of Nigeria, Nsukka.
4. Food and Agriculture Organization (FAO) 2003. The One to Watch, Radio, New ICTs and Interactivity. Rome: FAO, 57: 105.
5. Ozowa, V.N., 1995. Information needs of small farmers in Africa: The Nigerian example.